\documentclass[smallextended]{svjour3}

\smartqed  
\usepackage{graphicx}

\usepackage{layout}
\usepackage{textcomp}
\usepackage[toc,page,titletoc]{appendix}

\usepackage{subfigure}
\usepackage{graphics}
\usepackage{footnote}
\usepackage{epstopdf}
\usepackage{setspace}
\usepackage{tabularx,ragged2e,booktabs}
\usepackage{array}
\usepackage{verbatim}
\usepackage{multirow}
\usepackage{hhline}
\usepackage{natbib}
\usepackage{amsmath}
\usepackage{amssymb}
\usepackage{amssymb}

\newcommand{\arcsec}{\mbox{\ensuremath{\hspace{5pt}\!\!^{\prime\prime}}}}
\makeatletter

\newcolumntype{C}[1]{>{\centering\arraybackslash}p{#1}}
\newcommand{\Rmnum}[1]{\expandafter\@slowromancap\romannumeral #1@}
\DeclareMathSymbol{\mrq}{\mathord}{operators}{`'}

\newcommand\Tstrut{\rule{0pt}{3.8ex}}
\newcommand\Bstrut{\rule[-0.9ex]{0pt}{0pt}}

\usepackage{fancyhdr}
\usepackage{perpage}
\begin{document}

\title{An investigation of the Eigenvalue Calibration Method (ECM) using GASP for non-imaging and imaging detectors 
}


\author{Gillian Kyne \and David Lara \and Gregg Hallinan \and Michael Redfern \and Andrew Shearer}


\institute{G. Kyne \at
              School of Physics\\ National University of Ireland, Galway\\ Ireland. \\
              \email{gilliankyne@gmail.com}           
           \and
           D. Lara \at
              Cambridge Consultants \\ Science Park\\ Milton Road\\ Cambridge CD4 0DW\\ UK.
	   \and
           G. Hallinan \at
              Division of Physics Mathematics, and Astronomy\\ California Institute of Technology\\ Pasadena\\ CA 91125\\ USA. \\
	  \and
           M. Redfern \at
              School of Physics\\ National University of Ireland, Galway\\ Ireland. \\
	  \and
           A. Shearer \at
              School of Physics\\ National University of Ireland, Galway\\ Ireland. \\
	      \email{andy.shearer@nuigalway.ie}
}


\maketitle

\begin{abstract}

Polarised light from astronomical targets can yield a wealth of information about their source radiation mechanisms, and about the geometry of the scattered light regions. Optical observations, of both the linear and circular polarisation components, have been impeded due to non-optimised instrumentation. The need for suitable observing conditions and the availability of luminous targets are also limiting factors. The science motivation of any instrument adds constraints to its operation such as high signal-to-noise (SNR) and detector readout speeds. These factors in particular lead to a wide range of sources that have yet to be observed. The Galway Astronomical Stokes Polarimeter (GASP) has been specifically designed to make observations of these sources.
\par
GASP uses division of amplitude polarimeter (DOAP) \citep{Compain1998} to measure the four components of the Stokes vector (I, Q, U and V) simultaneously, which eliminates the constraints placed upon the need for moving parts during observation, and offers a real-time complete measurement of polarisation. Results from the GASP calibration are presented in this work for both a 1D detector system, and a pixel-by-pixel analysis on a 2D detector system. 
\par

Following \cite{Compain1999} we use the Eigenvalue Calibration Method (ECM) to measure the polarimetric limitations of the instrument for each of the two systems. Consequently, the ECM is able to compensate for systematic errors introduced by the calibration optics, and it also accounts for all optical elements of the polarimeter in the output. Initial laboratory results of the ECM are presented, using APD detectors, where errors of 0.2\% and 0.1$^{\circ}$ were measured for the degree of linear polarisation (DOLP) and polarisation angle (PA) respectively. Channel-to-channel image registration is an important aspect of 2-D polarimetry. We present our calibration results of the measured Mueller matrix of each sample, used by the ECM, when 2 Andor iXon Ultra 897 detectors were loaned to the project. A set of Zenith flat-field images were recorded during an observing campaign at the Palomar 200 inch telescope in November 2012. From these we show the polarimetric errors from the spatial polarimetry indicating both the stability and absolute accuracy of GASP.

\end{abstract}

\section{Introduction}
\label{Introduction}

A Stokes polarimeter is described as \textquoteleft{complete\textquoteright} when the polarisation state analyser (PSA) can measure the four linearly independent states of polarisation of light. The calibration of a polarimetric tool is a two-step process. Firstly, the optical elements of the system must be precisely orientated with very careful alignment, and secondly, the behaviour of these optical elements must be easily described and characterised by physical properties \citep{Compain1999}. Therefore, the combination of several measurements in different configurations, to determine the main values and shortcomings. Any optical element in the light path that has not been fully characterised, can introduce modifications to the state of polarisation of light and bias the estimation of the Stokes parameters.
\par
The linear component of optical polarised light emission has been well documented, however, observations have been limited due to instrumental considerations to periods of excellent observing conditions, and to steady (or slowly or periodically-varying) sources \citep{Collins2013}. In astronomical polarimetry, the optical circular component is less established. One important consideration in the development of any instrument, particularly in this work, is the ability to achieve adequate SNR for the science action, or minimise all sources of instrumental error.
\par
GASP was designed as a DOAP polarimeter using a modified Fresnel rhomb \citep{Compain1998}. \cite{Collins2013} reported an initial design of GASP and results based on a pseudo-inverse calibration. In order to calibrate a polarimeter, the polarimetric optics/materials calibrating the polarimeters must also be calibrated. The orientation of the elements that make up the polarisation state generator (PSG) - the angle of transmission axis of the polariser and the fast axis of the waveplate - must be calculated. If there are any errors in the PSG, used to calibrate/train the PSA, these errors will also propagate through to the instrument system matrix. 
\par
The ECM has the ability to calibrate a Stokes polarimeter in the exact configuration in which it will be used. It is also effective in the case where the PSG is a required output of the calibration process, which can be found in a number of references including \cite{Compain1999,deMartino}. Another advantage of this method is that it uses simple, off the shelf glass optics that are characterised, and calibrated, during the ECM. The ECM can independently calibrate the PSA and PSG, and works in such a way that the instrument matrix can never be polluted by errors (described above) from the PSG. The main science motivation for this instrument is time resolved observations of the Crab pulsar, which has a period of approximately 33 ms and magnitude of only 16.5. These characteristics require the removal of all/any instrumental error, including that of calibration.
\par
We report the operation of GASP using an ECM approach, using two different optical-detector systems. The behaviour of any optical or electronic device can be distorted by first-order failures, as in the case of a photo-elastic modulator \citep{Drevillon1993}. Therefore, several measurements must be combined in a number of different configurations to determine the main values and the shortcomings. The optical set-up is then adapted to account for these measurements. These aspects of the calibration process have already been reviewed in the literature \citep{Hauge1978, Thompson1980, Azzam1989}. Furthermore, one can notice that usual calibration procedures are difficult to implement \textit{in-situ} because of the influence of the optical elements that are necessarily included in the light path (filters, windows, lenses, mirrors, etc.) All these elements can induce modifications of the state of polarisation of light.
\par
It has been discussed that the ECM minimises the main sources of error generally associated with calibration where it requires a complete system (the polarimeter), a Polarisation State Generator (PSG) and 4 samples. No assumptions are made to the system, the PSG and Mueller matrices of each sample are actually characterised by the ECM. The system matrix is simply a function of the wavelength and must be consistent from calibration to science.
\par
The general form for the matrices A and W are n $\times$ 4 and 4 $\times$ m, respectively, where \textit{n} and \textit{m} are greater than 4. \textit{n} {\textgreater} 4 mean that more than four input Stokes vectors are generated, and \textit{m} {\textgreater} 4 means that the output Stokes vectors are over-determined. This is the case for GASP. The ECM is fully compatible with other size matrices and the dimensions of the intermediate matrices must be adapted \citep{Compain1999}.
\par
An investigation of the ECM, using the two configurations of GASP, one using Avalanche Photodiodes (APDs), and one using imaging detectors, will be presented in two stages. First, the APDs were used to record data from a laboratory experiment. Then 2D ECM data was obtained during a 4 night observing run at the 200 inch Hale telescope at Palomar Observatory. The project was loaned 2 Andor iXon Ultra 897 EMCCD (Electron Multiplying Charge Coupled Device) detectors for the 4 night campaign. As well as an experimental calibration run, GASP was also used to observe the Crab pulsar (among other targets) at a frame rate $>$ 1000 Hz to obtain a phase-resolved polarimetric signal.

\section{The Galway Astronomical Stokes Polarimeter}
\label{GASP}

\begin{figure}[h]
\centering
\includegraphics[scale=0.4]{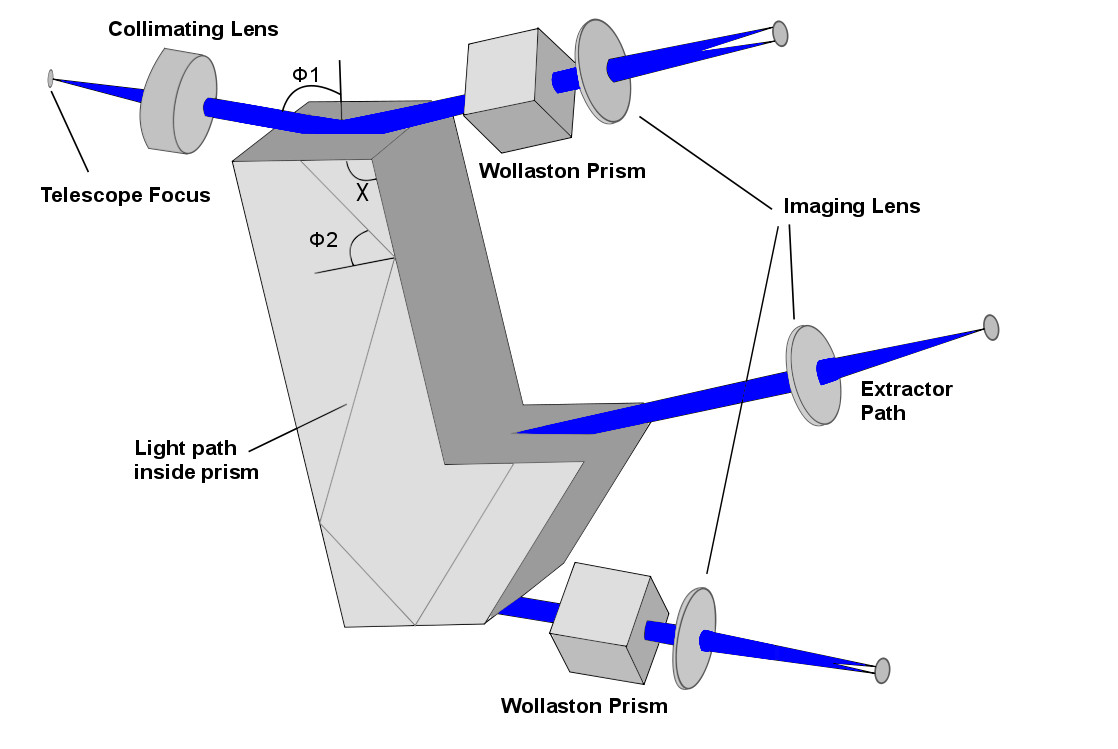}
\caption{The optical layout for a 2D detector system using GASP; this is a general design that can be adapted for a specific telescope. The reflected path (RP) is found on the top right hand corner, and the transmitted path (TP) is found at the bottom right hand corner. The optical prescription (with the exception of the main beam splitting prism) changes depending on the choice of telescope and detector.}
\label{GASP_Layout}
\end{figure}

GASP uses a modified DOAP design, which was implemented in the laboratory by \cite{Compain1998}. A full description of the polarimeter, in its initial design stage, is found in \cite{Collins2013}, with further optical development detailed in \cite{Kyne_thesis2014}. Figure \ref{GASP_Layout} gives a general description of the polarimeter 

\subsection{The Polarisation State Analyser}
\label{PSA_section}

The Polarisation State Analyser, PSA, is the instrument used to estimate the Stokes vector of a beam of light. In the case of an imaging polarimeter, the PSA is used to estimate a separate Stokes vector for each field position in the image. In GASP, the PSA is the combination of the main beamsplitting Fresnel rhomb, two polarising beamsplitters such as Wollaston or Foster prisms, the four simultaneous detectors, and any optical elements between the prism and the detectors. The PSA in GASP was originally designed as an 8-detector/1D system, using Foster prisms, where FOV size and image quality were not of great importance. The Wollaston (Foster) prisms must be aligned to an integer multiple of 45$^{\circ}$, with respect to the optical axis of the instrument. A 45$^{\circ}$ angle produces an invertible calibration matrix and splits each beam 50:50, once the system has been aligned experimentally; it must be calibrated based on this alignment. The PSA, with its base in the plane of incidence, acts as a first-stage beamsplitter and provides the essential polarimetric properties. These properties are displayed, as usual, by the reflection and transmission Fresnel coefficients of the prism \citep{Azzam1988Book},
\par
The PSA has a characteristic matrix, A, that fully describes the instrument polarimetrically, i.e., it contains all the information for how well the instrument is calibrated, experimentally. This matrix is the output of the ECM and contains both the system and demodulation matrix, which can be described best as a theoretically corrected system matrix, which is a better representation of the hardware used in constructing GASP. Another inclusion to the A matrix is the detector gain; the ECM allows for a different gain on each detector/pixel. The final Stokes vector S, is given by the following relationship:

\begin{equation}
 \rm{S = A^{-1}I}
\label{SAI}
\end{equation}

where A is the $4 \times 4$ experimental system matrix and I is the $4 \times n$ intensity array. n can be of any length depending on the acquisition time of the data. S will be the same size as I.
\par
\cite{Collins2009} describes a design where the four components of the Stokes vector are measured simultaneously with a retardance error $\ll$1\%, over a spectral range of 400 - 800 nm. The DOAP configuration contains no moving or modulating components, where it is possible to achieve high time resolution (of order microseconds), and temporal stability based on the choice of detector. The optical configuration of GASP can vary for reasons explained in \cite{Kyne_thesis2014}, depending on instrument size and detector availability. The values in Table \ref{tab:prism_rhomb_table} are the properties corresponding to the current GASP rhomb and configuration for the instrument which observed at Palomar. All measurements, theoretical and experimental, use these values.

\begin{table}[h]
\caption{Geometrical and polarimetric properties of the current GASP prism-rhomb from \cite{Collins2013} and \cite{Kyne_thesis2014}.}
\label{tab:prism_rhomb_table}
\resizebox{1.0\columnwidth}{!}{%
\begin{tabular}{llll}
\hline\noalign{\smallskip}
Prism Geometry & Glass Properties & \multicolumn{2}{l}{Polarimetric Properties} \\
\noalign{\smallskip}\hline\noalign{\smallskip}
\multicolumn{2}{l}{} & Reflection & Transmission\\
\noalign{\smallskip}\hhline{~~--}\noalign{\smallskip}
$\phi$1 = 78.50$^{\circ}$ & $\lambda$ = 589nm & R = 0.3533 & T = 0.4352 \\\Tstrut\Bstrut $\phi$2 = 58.48$^{\circ}$ & n = 1.589 & $\Delta_{r}$ = 180$^{\circ}$ & $\Delta_{t}$ = 90$^{\circ}$\\
\Tstrut\Bstrut $\chi$ = 96.62$^{\circ}$ & $\alpha$ = 0.2002 & $\Psi_{r}$ = 30.48$^{\circ}$ & $\Psi_{t}$ = 59.85$^{\circ}$\\
\noalign{\smallskip}\hline
\end{tabular}}
\end{table}

\section{The Eigenvalue Calibration Method}
\label{mathsECM}

The ECM sets out to eliminate/minimise the main sources of error generally associated with calibration. Three characteristic matrices are described, W for the entrance arm (PSG), A for the polarimeter, and \textbf{M}, the Mueller matrix of the sample. The GASP PSG has been modified to that described by \cite{Compain1999} and has been discussed by \cite{Kyne_thesis2014}. A measurement of 

\begin{equation}
 \rm{I = A\mathbf{M}W}
\label{AMW}
\end{equation}

is the starting point of the ECM, making use of linear algebra. The ECM has the following advantages \citep{Compain1999}:

\begin{enumerate}
 \item No assumption is made to the system to be calibrated - except that it must be complete. The precise orientation and position of the various elements that comprise the polarimeter do not need to be known for calibration. All optical elements are included in the matrix representation.
 \item The characteristics of the reference samples are completely determined during the calibration without need for secondary measurements. All aspects of the sample are measured during the ECM. W and A, when measured, will be a function of the wavelength $\lambda$, as will the defining characteristics of the reference samples. 
 \item The accuracy of the calibration procedure is evaluated when the ECM is used.
\end{enumerate}

Therefore, according to the above points, the Stokes vectors for the PSG used by GASP does not need to be known prior to calibrating the instrument. It is in fact calculated during the calibration process using Equation \ref{AMW}. The following description is based on the work by \citep{Compain1999} and references methods and naming conventions used by \citep{deMartino} and \citep{David_Lara_thesis}. A more detailed description of the mathematical solution for this method is found in the above references, including a description of the PSG used for calibration in \cite{Kyne_thesis2014}.
\par
A linear mapping is described by

\begin{align}
 \mathbb{H}_{4}:\mathbb{M}(\mathbb{R}) &  \to \mathbb{M}_{4}(\mathbb{R}), \\
 X & \to \textbf{M}X - X(\textit{aw})^{-1}(\textit{amw}). 
 \end{align}

\label{ECM_mapping}

\textbf{M} is the Mueller matrix of the reference sample and (\textit{amw}) corresponds to experimental measurements. $\mathbb{H}$ has the property of having W within its null space, i.e.,

\begin{equation}
 \mathbb{H}\rm{(W) = 0},
\label{HW=0}
\end{equation}

whatever the value of \textbf{M}, because without experimental errors $(\mathit{aw})^{-1}(\mathit{amw})$ is equal to $\rm{W}^{-1}\mathbf{\rm{M}}\rm{W}$. A well chosen set of reference samples for $\{\mathbf{\rm{M}}_{i}....\mathbf{\rm{M}}_{n}\}$ can reduce the number of solutions to Equation \ref{HW=0} to one. And then A can be deduced from the following:

\begin{equation}
 \rm{A} = (\mathit{aw})W^{-1}.
\label{A=awW}
\end{equation}

The chosen reference samples for GASP are based on the work carried out by \cite{deMartino} and \cite{David_Lara_thesis} and are discussed in detail by these references.

\section{GASP ECM: Methods}
\label{methodsECM}

A number of calibration data sets were recorded, while GASP was mounted at the Cassegrain focus of the 200 inch Hale telescope at Palomar Observatory. An optical lens prescription was used to change the nominal f/16 beam to f/9, providing a more acceptable FOV to the GASP PSA. The calibration images are acquired as follows:

\begin{itemize}

\item Input light is generated using an LED, a filter, and a linear polariser.

\item An ECM wheel contains a set of samples, AIR, a polariser at 0$^{\circ}$, a polariser at 90$^{\circ}$ and a quarter waveplate at 30$^{\circ}$.

\item The polarimeter output intensities are recorded for a set of angles ranging from -38$^{\circ}$ to -209.36$^{\circ}$ in -10.08$^{\circ}$ increments for each of the above samples, and is repeated using a quarter waveplate (QWP) placed in front of the linear polariser. Though of course, as has been explained above, it is not necessary to know what these angle are, as long as the same process is repeated for the QWP.

\item These intensities are used to extract a system matrix, A and Polarisation State Generator (PSG), W.

\end{itemize}

The system matrix, A, is normalised to the first element of the theoretical GASP PSA using Equation \ref{A_norm} to determine variations in the 16 coefficients, and a comparison between the individual coefficients. As the experimental matrix is calculated based on intensity values, it is difficult to interpret the variation in the coefficients without normalisation.

\begin{equation}
\rm{A_{norm,i} = \frac{A_{i}PSA(1,1)}{A_{i}(1,1)}}.
\label{A_norm}
\end{equation}

\subsection{GASP ECM: APDs Results}
\label{ECM_results_APDs}

The ECM is demonstrated in the laboratory using Avalanche Photodiodes, where LED light is fed to each detectors, through the polarimeter, by optical fibres. A description of the optical setup for these detectors is found in \cite{Kyne_thesis2014}. 
A set of calibration data was recorded, using APDs, in the laboratory dated July 2011 and an experimental system matrix, A, shown by,

\begin{equation}
\rm{A} = \left( \begin{array}{cccc} 0.1766 & -0.0901 & 0.1571 & 0.0042\\ 0.4983 & -0.2010 & -0.4419 & 0.0113\\ 0.4992 & 0.2503 & 0.1167 & -0.4058\\ 0.3813 & 0.1826 & -0.0535 & 0.3229\end{array} \right).
\label{A_apds}
\end{equation}

which is compared to the theoretical PSA,

\begin{equation}
\rm{PSA} = \left( \begin{array}{cccc} 0.1766 & -0.0858 & 0.1544 & 0\\ 0.1766 & -0.0858 & -0.1544 & 0\\ 0.2176 & 0.1078 & 0.0008 & -0.1890\\ 0.2176 & 0.1078 & -0.0008 & 0.1890\end{array} \right).
\label{PSA2}
\end{equation}

In the absence of errors, and optical configuration differences, these matrices should match. The PSA matrix has been theoretically corrected to produce the system matrix A, which includes polarimetric errors and offsets from the design. The matrix is divided into 4 rows. The first 2 rows represent the RP1 and RP2 channels, and the second 2 rows the TP1 and TP2 channels. The order of these channels depend on the optical alignment in the laboratory. The first column of the system matrix (and the PSA) represents the gain of the 4 channels of the DOAP, which includes the gain of each detector. The detector gain of each APD accounts for the main difference in the overall PSA gain coefficients.
\par
The system matrix A above, was used on a set of data set to reduce the Stokes parameters for a polarised light source using an LED, and a glass polariser (Sodium-Silicate glass). The data was recorded for approximately 48 hours. The temporal stability of the GASP detectors was not under investigation. A number of APDs were deemed unsuitable in terms of dark counts and electronic noise after these laboratory experiments.

\begin{figure}[h]
  \centering
  \includegraphics[width=1.0\linewidth,height=0.6\linewidth]{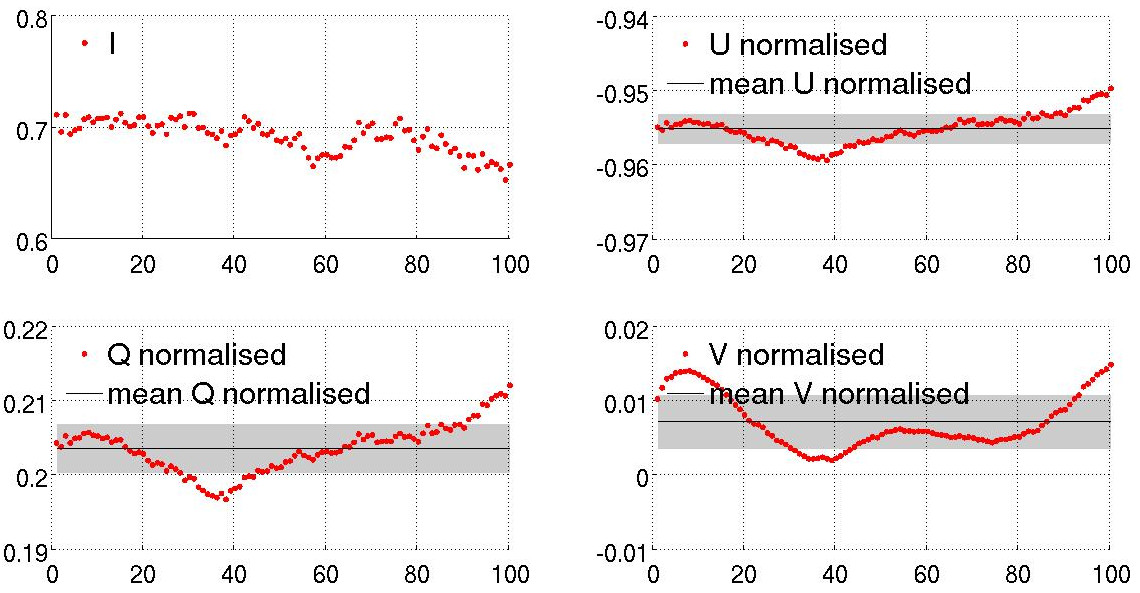}
  \caption{Stokes parameters I, Q, U, and V for an experiment passing LED light through a linear polariser to measure the limits of the GASP system using APDs. Each point is an average measurement of 1750 seconds of data. The Q, U, and V parameters have been normalised to Stokes I to remove any fluctuations in luminosity. A systematic effect is observed on Stokes Q and U, the parameters used to the measurement of the degree of linear polarisation. A different trend is found for V, which suggests an instrumental effect for this laboratory experiment.}
  \label{20110722_LP_norm_Stokes_errorpatch_crop}
\end{figure}

\begin{figure}[h]
  \centering
  \includegraphics[width=1.0\linewidth,height=0.6\linewidth]{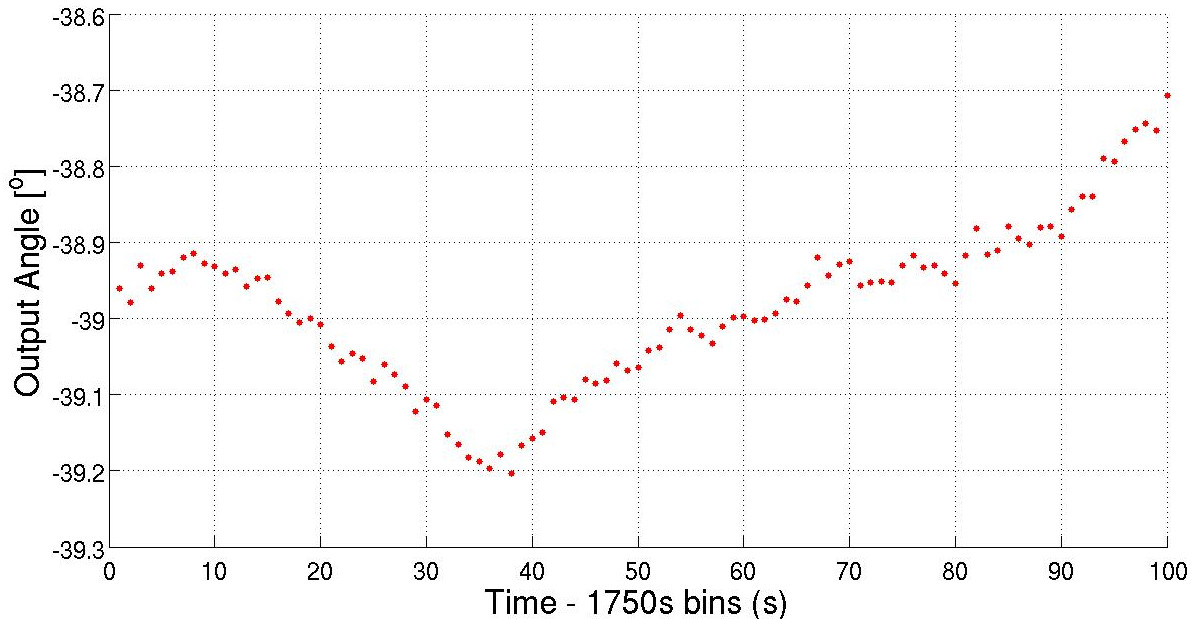}
  \caption{Polarisation angle measurement for an experiment passing LED light through a linear polariser to measure the limits of the GASP system using APDs. The drop, and variation, in Stokes Q and U is observed in this plot. There is a variation of 0.5$^{\circ}$ from beginning to end, which suggests a number of errors including thermal or pressure changes, or that the polariser moved during the experiment. This is not a random effect and over a prolonged period (2 days) it is possible that there was movement in the laboratory, or in the vicinity of the instrument to account for this variation.}
  \label{20110722_LP_outputangle_new_crop}
\end{figure}

The parameters Q, U, and V have been normalised to Stokes I. A trend in I indicates a drop in light intensity, possibly due to an intensity variation in the LED used over the 48 hour period. A change in the laboratory (or LED) temperature could also cause this effect, or vibrations due to any movement in the laboratory. After normalisation, Q, U, and V show evidence of systematic (including random) error over this period. Q, U, and V give mean values of 0.212 $\pm$ 0.003, -0.950 $\pm$ 0.002, and 0.015 $\pm$ 0.004, respectively. The standard deviation value is quoted after normalisation, however, there is a visible trend in each of these parameters that suggests a systematic change in the data. This is a possibility as during the experiment a linear polariser was used to generate a polarised signal. The systematic effect shows a drop in value at time point 38 (18.5 hours into the experiment) for all Q, U, and V. If the polariser moved, and the movement was not symmetric, then this could induce (or reduce) the measured linear and circular polarisation. It would also account for the change in polarisation angle seen in Figure \ref{20110722_LP_outputangle_new_crop}. This systematic error means that the errors quoted for this data set are higher than the random noise, which is visible in this plot.
\par
Over the time series the mean degree of linear polarisation (DOLP) was measured to be 97.66 $\pm$ 0.21\%, and the degree of circular polarisation (DOCP) 0.72 $\pm$ 0.36\%. This is a promising result that indicates GASP has the ability to measure small variations in the linear polarisation error as it fluctuates by about 0.2\%. The errors quoted are from a combined systematic and random effect, however, a polarimetric signal detecting small fluctuations has been measured.

The polarisation angle (PA) was measured prior to the laboratory experiment; this had a value of $\sim$ -38.5$^{\circ}$. Figure \ref{20110722_LP_outputangle_new_crop} plots a mean angle of -38.99 $\pm$ 0.19$^{\circ}$ measured by the ECM with an error level of 0.2$^{\circ}$.
\par
There is variation over the time series ranging over 0.5$^{\circ}$. The laboratory experiment contains about 48 hours of data in an uncontrolled environment subject to temporal fluctuations and potential mechanical disturbances. The temperature was found to be variable over the course of the experiment, but this data was not recorded. It is also noted that structure holding the linear polariser was subject to some movement within the mounting device, but it is not possible to prove that this is what caused the change in PA. If the variation in the PA is examined on a shorter time scale - in an area where the systematic effect is reduced, or absent - the variation is less than 0.1$^{\circ}$. This is a good indication that GASP is working correctly, and it also illustrates the capability of GASP to detector fluctuations in the PA to a level of $\pm$ 0.1$^{\circ}$.

\subsection{GASP ECM: EMCCDs Results}
\label{ECM_results_EMCCDs}

There were obstacles encountered using the APDs from Section \ref{ECM_results_APDs}, which included failing modules and high noise levels and dark counts. 2 Andor Ultra 897 EMCCD detectors were loaded to the project by a collaborating group at the California Institute of Technology, therefore, GASP was modified for use as an imaging polarimeter.
\par
A number of observation targets were used to produce a geometrical transformation of the RP1 channel with each of the RP2, TP1, and TP2 channels \citep{Kyne_thesis2014}. \cite{Tyo2006review} and \cite{Smith1999} discuss the critical issue of image registration for sequential or simultaneous image collection. Registration to ${1}/20$ of a pixel can achieve less than 0.01 error in DOLP and DOCP results. Misregistration can result, for example, from separate focal planes that are not looking at the same region of space; it can also result in beam wander from a rotating element. In practice, achieving even a half-pixel alignment mechanically is difficult - of course this will be dependent on the pixel size of the detector - and software post-processing alignment is frequently necessary. 
\par
GASP overcomes the problem of motion of the target field across the FOV using the DOAP technique. Multiple images per channel are acquired simultaneously, removing the need for continuous guiding. However, spatial registration of multiple images is complicated by the need to correct for both mechanical misalignment as well as optical \textquoteleft{misalignment\textquoteright} arising from aberrations due to separate optical paths. The TP has a longer optical path length compared to RP, and it was found in \cite{Collins2009} that a design feature of the transmitted path resulted in an astigmatism after transmission and before reimaging. It was not possible to correct this aberration for the purpose of these observations.

\par
Registration is carried out using matched points, channel-to-channel, from a set of targets on Night 4 of the Palomar observing run. \cite{Kyne_thesis2014} gives a detailed explanation for the point selection process using observational data. The registration routine involved the following steps:

\begin{enumerate}
 \item A reference channel is selected - RP1 is chosen for this data analysis.
 \item A set of points are matched in a region of interest (ROI) to be calibrated (analysed) between RP1 and the remaining channels.
 \item \textit{geomap}, an IRAF command, is used to generate a geometric transformation between RP1 and the remaining channels based on this matching.
 \item \textit{gregister} (from IRAF) is then used to register the images.
\end{enumerate}

\begin{figure}[h]
  \begin{minipage}[b]{0.45\linewidth}
    \centering
     \includegraphics[scale=0.55]{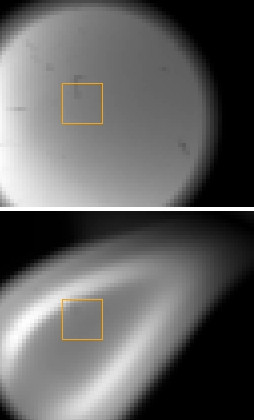}
  \end{minipage}
  \hspace{0cm}
  \begin{minipage}[b]{0.45\linewidth}
    \centering
     \includegraphics[scale=0.55]{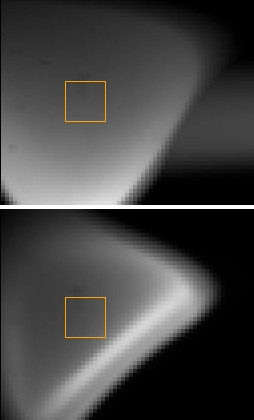}
  \end{minipage}
  \caption{All registered GASP channels using science data points and image artefacts. A pixel-by-pixel calibration analysis will be analysed on the area indicated by the orange box. The shape of the RP2, TP1, and TP2 images are not perfectly matched to RP1 due to insufficient matched points. IT is also noted that the optical aberration on the TP is too severe to correct in software alone. Top LHS: RP1, top RHS: RP2, bottom LHS: TP1 and bottom RHS: TP2.}
  \label{ALL_flats_science_moved_reg}
\end{figure}

The central region of the images in Figure \ref{ALL_flats_science_moved_reg} appear to be well registered, however, looking from the centre out to the edges of the field the matching deteriorates. This is expected given the limited number of matched points, and overall optical image quality. A pixel-by-pixel polarimetric analysis was carried out on an area of the FOV that showed minimum distortion.


\begin{figure}[h]
  \centering
  \includegraphics[scale=0.32]{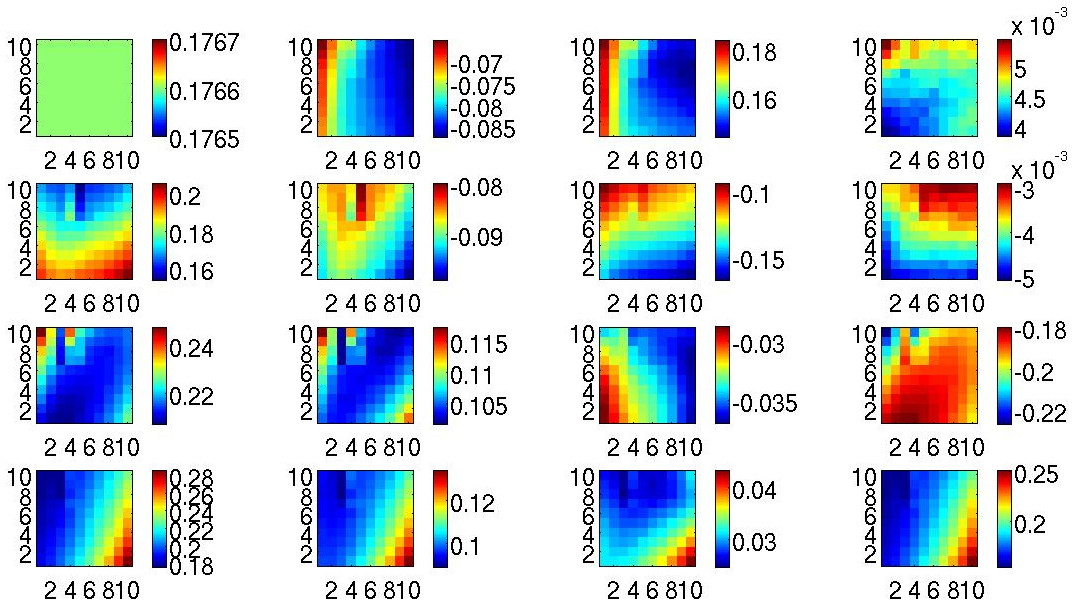}
  \caption{A pixel-by-pixel experimental calibration matrix for the image section ($\sim$ 10\arcsec) indicated by the orange box in Figure \ref{ALL_flats_science_moved_reg}. The calibration images were registered to channel RP1 using science data.}
  \label{A_norm_flat_moved_section_4x4_science_crop}
\end{figure}


Figure \ref{A_norm_flat_moved_section_4x4_science_crop} is an image of the normalised calibration system matrix generated from this registration solution. Each pixel has its own coefficient which forms a separate system matrix, for all pixels in this area of the FOV. In this way the analysis can be compared to that used for APDs, as each pixel is be thought of as a single APD. The main difference in this calibration analysis is that the result is field dependent, as when the pupil is reimaged it is spread across a number of pixels, instead of a single APD.
\par
It has been discussed that the first column of the matrix gives the gain for each channel. The first row is a measure for RP1, row 2 - 4, are RP2, TP1, and TP2, respectively. There is a variation (banding) in the gain values, showing a systematic trend across each coefficient in column 1, with the exception of gain coefficient RP1. On examination of the trends found in the image residuals after registration \citep{Kyne_thesis2014}, the patterns observed are nearly identical. It was not possible to completely eliminate all optical distortion using post-processing techniques, resulting in these systematic effects. Normalisation gives a good indication of how well registered the RP2, TP1, and TP2 channels are compared to RP1. The system matrix is used to perform the Stokes reduction for science data in Section \ref{zenith_obs}. 
\par
The gain values for each pixel for RP1 and RP2 compare well, as do those for TP1 and TP2 getting worse towards the edge of the field. These values are compared to the theoretical PSA, though, it is not absolutely correct to compare this to the theoretical PSA given the differences that will exist in optical alignment. In the absence of variation in luminosity in the field, of each channel, a more random pattern would be observed in each coefficient. Where the values are expected to be low ($\sim$ 0), a pattern will be less evident where the random errors are approximately equal to the systematic, taking into account alignment variation in the theoretical PSA.

\subsection{GASP ECM: Verification}
\label{verifyECM}

It is important that the results of the ECM are verified experimentally to ensure that the system is calibrated. This is done by testing the system matrix with another set of calibration data, which is passed through the same samples used for the ECM. A number of parameters are checked in this way: the polarimetric offsets/error as a function of pixel, or subsection of the FOV, and the field-dependence. Each ECM sample has been aligned to a particular value, which the ECM measures as an output of the calibration procedure. There is optical field distortion present in the system, which could not be completely removed in software; we can map the calibration, pixel-to-pixel variation over the FOV by this verification process.
\par
\begin{figure}[h]
  \centering
  \includegraphics[scale=0.32]{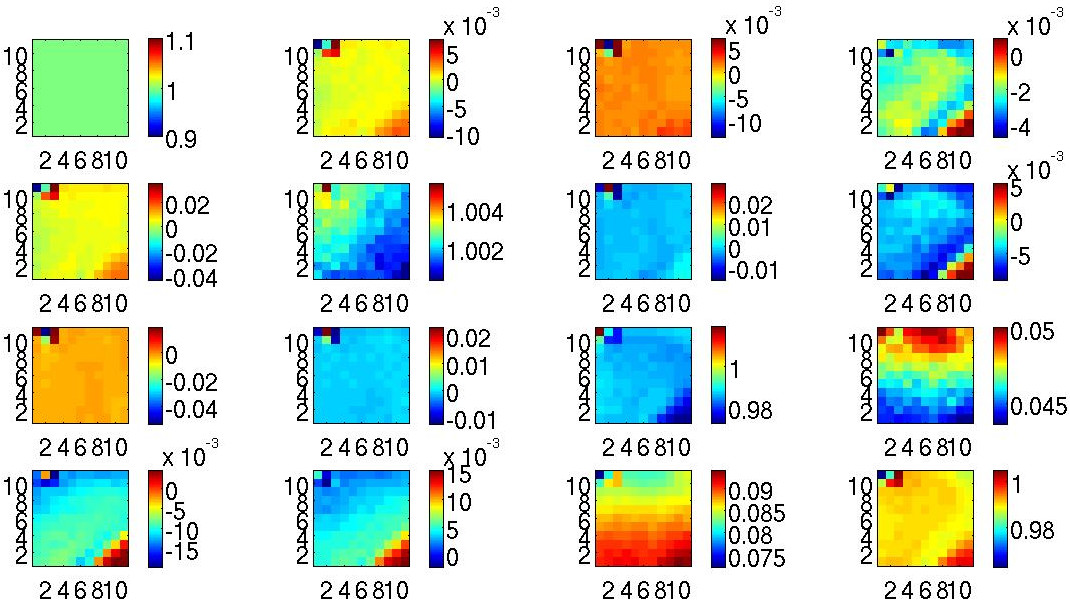}
  \caption{The normalised Mueller matrix for the AIR sample for the image section indicated in Figure \ref{ALL_flats_science_moved_reg}. The calibration images were registered to channel RP1 using science data.}
  \label{AIR_science_crop}
\end{figure}

A systematic trend can be found across the images of the coefficients in Figure \ref{A_norm_flat_moved_section_4x4_science_crop}, which correlates with the patterns visible in the raw data after registration. The Mueller matrix for the AIR ECM sample is found in Figure \ref{AIR_science_crop}. The $\sigma$ values are as low as 0.5\%, compared with the expected MM. There is evidence of a systematic effect on each of the TP images, particularly TP2. This is consistent with the fact that there is more observed distortion on this path. The theoretical and mean value for the second to last coefficient is quite high; this is indicative of the difficulty in image registration. The same is found for the last TP1 coefficient.


\begin{table}[h]
\begin{center}
\caption{Theoretical and mean values for the normalised Mueller matrix of the AIR ECM sample; registration performed using science data points.}\label{tab:mean_theo_AIR_sample}
\resizebox{1.03\columnwidth}{!}{%
\begin{tabular}{c c}
\hline\noalign{\smallskip}
$\rm{\mathbf{MM_{AIR}}}$ & $\mathbf{\bar{\rm{x}}}$ \\
\noalign{\smallskip}\hline\noalign{\smallskip}

$\left(\begin{array}{cccc} 1.0000 & 0.0000 & 0.0000 & 0.0000\\ 0.0000 & 1.0000 & 0.0000 & 0.0000\\  0.0000 & 0.0000 & 1.0000 & 0.0000\\ 0.0000 & 0.0000 & 0.0000 & 1.0000\end{array} \right)$


&

$\left(\begin{array}{cccc} 1.0000 & 0.0004 & 0.0014 & -0.0020\\ 0.0043 & 1.0021 & -0.0018 & -0.0045\\ -0.0032 & -0.0008 & 0.9854 & 0.0467\\ -0.0094 & 0.0046 & 0.0873 & 0.9902\end{array} \right)$ \\

\hline\noalign{\smallskip}
\end{tabular} }
\end{center}
\end{table}

\begin{figure}[h]
  \centering
  \includegraphics[scale=0.32]{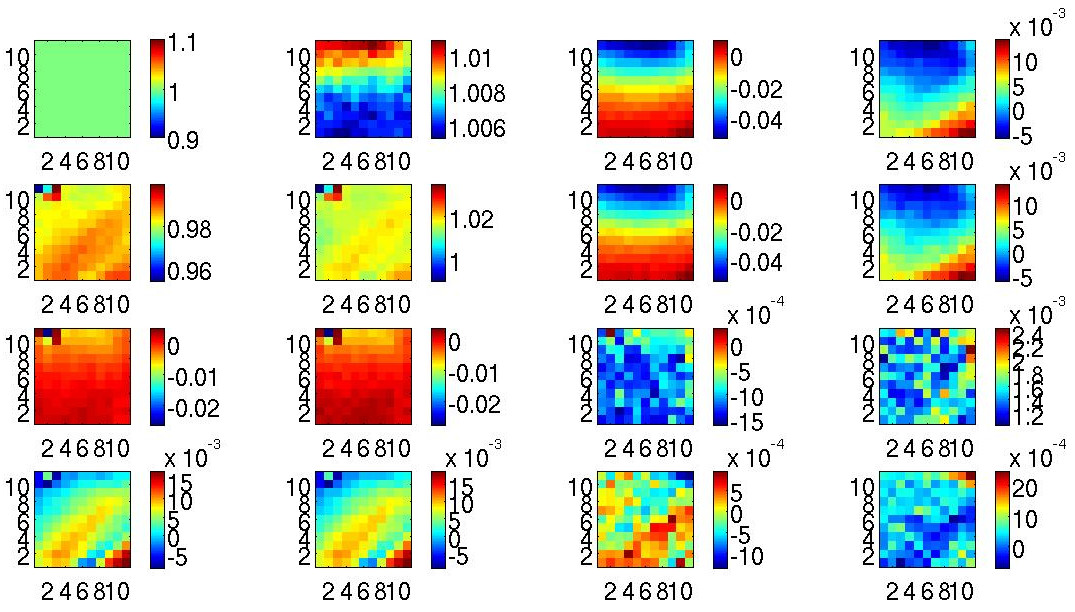}
  \caption{The normalised Mueller matrix for the P0 sample for the image section indicated in Figure \ref{ALL_flats_science_moved_reg}. The calibration images were registered to channel RP1 using science data.}
  \label{P0_science_crop}
\end{figure}

\begin{figure}[h]
  \centering
  \includegraphics[scale=0.32]{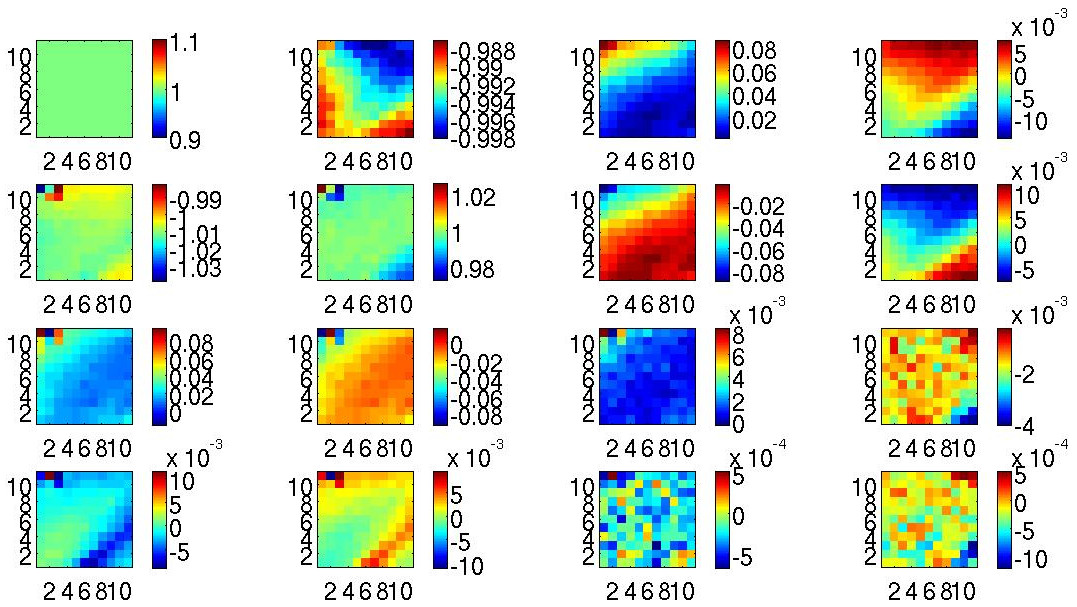}
  \caption{The normalised Mueller matrix for the P90 sample for the image section indicated in Figure \ref{ALL_flats_science_moved_reg}. The calibration images were registered to channel RP1 using science data.}
  \label{P90_science_crop}
\end{figure}

Figure \ref{P0_science_crop} and \ref{P90_science_crop} are the Mueller matrices for the linear polariser samples P0 and P90, respectively. These samples have similar MM, with the exception of sign. The variation is low on each channel, however, the values for the 3$^{rd}$ coefficients on rows 1 and 2 are higher than expected. This could be for a number of reasons, image registration, imperfections in the polarisers used, the polarisers may not exactly align to the stated value. The pixel values across the FOV will change as a function of angle of incidence between the sample and the detector - which is related to the quality of the reimaging optics.
\par
There is a trend on each of the TP rows (3 and 4). The top left and bottom right corner show larger variation compared to that in the centre of the FOV. It is most likely due to an absence of registration points in those areas. The P0 and P90 experimentally determined Mueller matrices show low variation with their corresponding theoretical matrices.

\begin{table}[h]
\begin{center}
\caption{Theoretical and mean values for the normalised Mueller matrix of the P0 ECM sample; registration performed using science data points.}\label{tab:science_mean_theo_P0_sample}
\resizebox{1.03\columnwidth}{!}{%
\begin{tabular}{c c}
\hline\noalign{\smallskip}
$\rm{\mathbf{MM_{P0}}}$ & $\mathbf{\bar{\rm{x}}}$ \\
\noalign{\smallskip}\hline\noalign{\smallskip}

$\left(\begin{array}{cccc} 1.0000 & 1.0000 & 0.0000 & 0.0000\\ 1.0000 & 1.0000 & 0.0000 & 0.0000\\  0.0000 & 0.0000 & 0.0000 & 0.0000\\ 0.0000 & 0.0000 & 0.0000 & 0.0000\end{array} \right)$


&

$\left(\begin{array}{cccc} 1.0000 & 1.0075 & -0.0166 & 0.0015\\ 0.9848 & 1.0169 & -0.0175 & 0.0017\\ -0.0004 & -0.0014 & -0.0011 & 0.0015\\ 0.0044 & 0.0049 & -0.0000 & 0.0003\end{array} \right)$ \\

\hline\noalign{\smallskip}
\end{tabular} }
\end{center}
\end{table}

The results found for the horizontal polariser in Figure \ref{P0_science_crop} are comparable to those in Figure \ref{P90_science_crop}, and the mean coefficient values can be found in Table \ref{tab:science_mean_theo_P0_sample} and \ref{tab:science_mean_theo_P90_sample}, respectively. The variation is similar and the proposed systematic effect of registration is also present in each of these images. These Mueller matrices show a variability compared to the field-dependence errors found in the calibration matrix (Figure \ref{A_norm_flat_moved_section_4x4_science_crop}).

\begin{table}[h]
\begin{center}
\caption{Theoretical and mean values for the normalised Mueller matrix of the P90 ECM sample; registration performed using science data points.}\label{tab:science_mean_theo_P90_sample}
\resizebox{1.03\columnwidth}{!}{%
\begin{tabular}{c c}
\hline\noalign{\smallskip}
$\rm{\mathbf{MM_{P90}}}$ & $\mathbf{\bar{\rm{x}}}$ \\
\noalign{\smallskip}\hline\noalign{\smallskip}

$\left(\begin{array}{cccc} 1.0000 & -1.0000 & 0.0000 & 0.0000\\ -1.0000 & 1.0000 & 0.0000 & 0.0000\\  0.0000 & 0.0000 & 0.0000 & 0.0000\\ 0.0000 & 0.0000 & 0.0000 & 0.0000\end{array} \right)$


&

$\left(\begin{array}{cccc} 1.0000 & -0.9930 & 0.0238 & -0.0002\\ -1.0091 & 0.9969 & -0.0242 & -0.0005\\ 0.0190 & -0.0181 & 0.0012 & -0.0017\\ -0.0017 & 0.0010 & -0.0002 & -0.0002\end{array} \right)$ \\

\hline\noalign{\smallskip}
\end{tabular}}
\end{center}
\end{table}

\begin{figure}[h]
  \centering
  \includegraphics[scale=0.32]{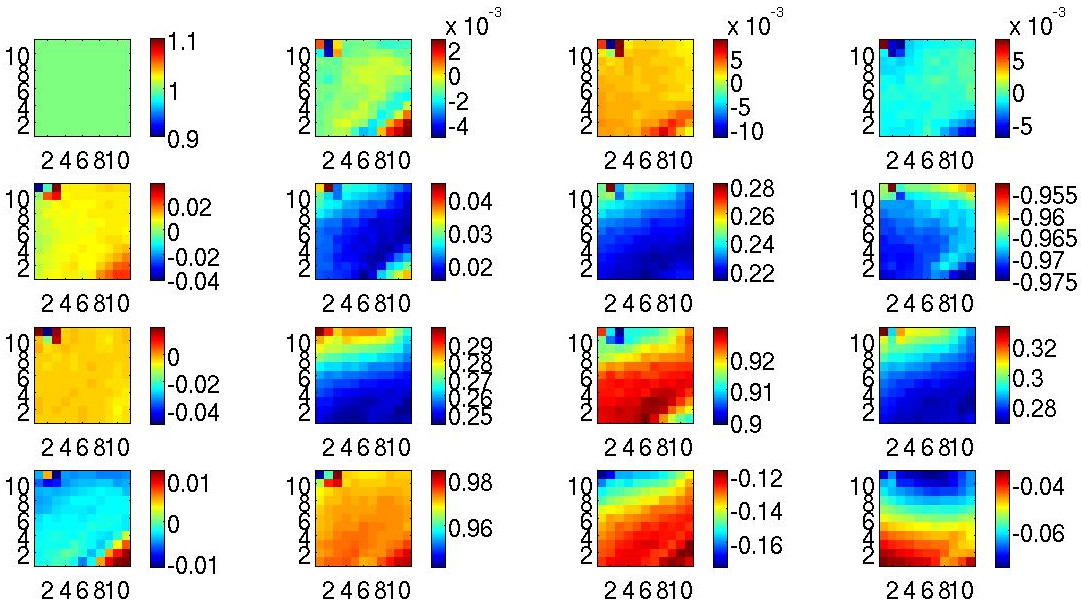}
  \caption{The normalised Mueller matrix for the R30 sample for the image section indicated in Figure \ref{ALL_flats_science_moved_reg}. The calibration images were registered to channel RP1 using science data.}
  \label{R30_science_crop}
\end{figure}

Figure \ref{R30_science_crop} is the Mueller matrix for the sample R30. It was concluded, from an analysis of averaged/binned data, that this sample has been measured to be closer to an angle of $\sim$ 38$^{\circ}$. The sample was originally aligned to approximately $\sim$ 30$^{\circ}$, a difficult mechanical task with the current setup, however, when calibration was carried out for a point source (a pinhole at the instrument focus, instead of an extended field) the ECM measured this sample to be $\sim$ 38$^{\circ}$, which is another very useful attribute of the method. Just like the figures above, there is a low systematic trend present which can be compared to that in the system matrix.
\par
Looking at Figure \ref{R30_science_crop}, the TP coefficients, rows 3 and 4, register the highest variation for all coefficients, about 1\%, which compares to the FOV variation for that of the system matrix (Figure \ref{A_norm_flat_moved_section_4x4_science_crop}). As mentioned above, this is most likely a result of incorrect pixel-matching.


\begin{table}[h]
\begin{center}
\caption{Theoretical and mean values for the normalised Mueller matrix of the R30 ECM sample; registration performed using science data points.}\label{tab:science_mean_theo_R30_sample}
\resizebox{1.03\columnwidth}{!}{%
\begin{tabular}{c c}
\hline\noalign{\smallskip}
$\rm{\mathbf{MM_{P38}}}$ & $\mathbf{\bar{\rm{x}}}$ \\
\noalign{\smallskip}\hline\noalign{\smallskip}


$\left(\begin{array}{cccc} 1.0000 & 0.0000 & 0.0000 & 0.0000\\ 0.0000 & 0.0585 & 0.2347 & -0.9703\\ 0.0000 & 0.2347 & 0.9415 & 0.2419\\ 0.0000 & 0.9703 & -0.2419 & 0.0000\end{array} \right)$



&

$\left(\begin{array}{cccc} 1.0000 & -0.0010 & 0.0023 & -0.0014\\ 0.0083 & 0.0212 & 0.2259 & -0.9689\\ -0.0040 & 0.2593 & 0.9217 & 0.2823\\ -0.0020 & 0.9720 & -0.1346 & -0.0545\end{array} \right)$ \\

\hline\noalign{\smallskip}
\end{tabular}}
\end{center}
\end{table}

The variation from the mean value for each coefficient of each Mueller matrix is low, and within the expected values for each of the RP1, RP2, TP1, and TP2 channels, which was discussed in Section \ref{ECM_results_EMCCDs}. A pattern remains in the FOV of a number of coefficients in each of the samples for AIR, P0, P90, and R30. The same pattern is found in Figure \ref{A_norm_flat_moved_section_4x4_science_crop} for the respective channels. The reason for this pattern can also be referenced in Figure \ref{ALL_flats_science_moved_reg} where there are insufficient points, and the areas absent of points are showing this trend.

\section{Zenith Flat-Field Observations}
\label{zenith_obs}

A 35 minute set of data was recorded at twilight on the last night of observations during the Palomar November observing run. Each frame was integrated for 1 second, continuously, as the Sun was rising. The data was recorded in this way to measure how the polarisation of the sky changes with respect to the elevation of the Sun over time, i.e. the polarisation of the sky is expected to increase as the elevation increases. This type of experiment gives valuable information about the accuracy of the GASP image registration/calibration by measuring the limit of the polarimetric variation over the FOV. It is also a very interesting experiment to determine a measurement of the degree of linear polarisation, and change in polarisation angle, of the twilit sky over time. 
\par
\cite{Dahlberg2009} and \cite{Harrington2011} describe how there are many atmospheric and geometric considerations, which determine the skylight polarisation at a particular observation site. The magnitude of the polarisation angle can depend on the solar elevation, atmospheric aerosol content, aerosol vertical distribution, aerosol scattering phase function, wavelength of the observation and secondary sources of illumination. Further descriptions of this can be found in these references \cite{Horvath2002,Suhai2004,Cronin2006}. Anisotropic scattered sunlight that can arise from reflections off land or water can be highly polarised and variable \cite{Peltoniemi2009,Litvinov2010,Kisselev2004}. Aerosol particle optical properties and vertical distributions in the atmosphere can vary and can cause variability on the incoming polarisation signal \citep{Shukurov2006,Vermeulen2000,Ugolnikov2004,Ugolnikov2010}. The polarisation can change across atmospheric absorption bands or can be influenced by other scattering mechanisms \cite{Boesche2006,Zeng2008,Aben1999,Aben2001}. The polarimetric measurements made by GASP \citep{Harrington2011}, will contain deviations from the Rayleigh Sky model, which will grow with aerosol, cloud, ground or sea-surface scattering, and affect the telescope line-of-sight. However, clear, cloudless, low-aerosol (ideal) conditions should yield high linear polarisation amplitudes and small deviations in polarisation, from the model \citep{Pust2005,Pust2006,Pust2007,Pust2008,Pust2009,Shaw2010}.
\par
This analysis will assess the performance of the GASP calibration, as well as carrying out scientific measurements. The Zenith data was recorded in 3 stages: The first was summed and averaged for 48 seconds, the first image in the figures to follow. Images 2 - 11 are averaged over 200 frames (3.3 minutes). The final image, the brightest sky image, is 134 seconds of data.

\subsection{Zenith Flat-Field Observations: Polarimetric Results}
\label{results_zenith_obs}

An approximation for the expected variation in polarisation angle related to time-per-frame is give in Table \ref{tab:pol_info_PA_expected}. The variation in DOLP is compared with that of the Rayleigh Sky Model, which gives the value for the degree of linear polarisation calculated as a function of the absolute angle for a reflected beam of light,

\begin{equation}
 \rm{DOLP}({\gamma}) \rm{=} \frac{1 - \cos^{2}\gamma}{1 + \cos^{2}\gamma}.
 \label{DOLP_angle_theo}
\end{equation}

$\gamma$ is the scattered angle formed between the (telescope) pointing and the Sun, and $\theta = 90^{\circ} - \gamma$, the angle made by the Sun, the pointing and the Zenith. According to Equation \ref{DOLP_angle_theo}, as the scattered angle increases, the DOLP also increases. A detailed experiment of how this model describes the E-vector was carried out by \cite{Suhai2004}.

\begin{savenotes}
\begin{table}[h]
\begin{center}
\caption{Expected polarisation angle variation over the same 35 minute period of observation by GASP for Zenith flat-field data. The first time is not usable as we are measuring from this value: The $\Delta$PA values are measured from here.}
\label{tab:pol_info_PA_expected}
\resizebox{1.0\columnwidth}{!}{%
\begin{tabular}{l l l l l l l l l l l l l}
\hline\noalign{\smallskip}
\textbf{Flat-field} & \textbf{1} & \textbf{2} & \textbf{3} & \textbf{4} & \textbf{5} & \textbf{6} & \textbf{7} & \textbf{8} & \textbf{9} & \textbf{10} & \textbf{11} & \textbf{12}\\ 
\noalign{\smallskip}\hline\noalign{\smallskip}
Time (mins) & 0.47 & 2.45 & 5.75 & 9.05 & 12.35 & 15.65 & 18.95 & 22.25 & 25.55 & 28.85 & 32.15 & 35.55 \\ 
\noalign{\smallskip}\hline\noalign{\smallskip}
$\Delta$PA ($^{\circ}$) & \centering{-} & 0.42 & 0.69 & 0.69 & 0.69 & 0.69 & 0.69 & 0.69 & 0.69 & 0.69 & 0.69 & 0.71 \\ 
\hline\noalign{\smallskip}
\end{tabular}}
\end{center}
\end{table}
\end{savenotes}

The values for time in Table \ref{tab:pol_info_PA_expected} were calculated from the time-stamps in the header (data information) files. The Sun moves 1$^{\circ}$ in 4 minutes and, given that the elevation at Palomar is +32.3$^{\circ}$, the expected change in PA is 0.84$^{\circ}$ in 4 minutes. The times given in Table \ref{tab:pol_info_PA_expected} are the midpoints of the recorded data - taking into account delays in acquisition due to stoppages.

\subsection{Discussion}

\begin{figure}[t]
  \centering
  \includegraphics[scale=0.58]{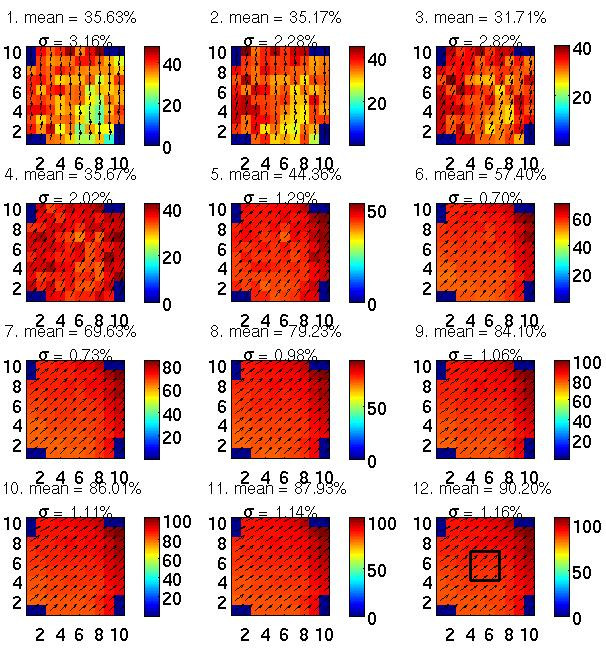}
  \caption{The degree of linear polarisation as a function of time, and field, are plotted. These images are produced using a registration method by science data points. The DOLP is measured as the Sun rises; the first was summed and averaged for 48 seconds, images 2 - 11 are averaged over 200 frames (3.3 minutes), and the final image, the brightest sky image, is 134 seconds of data. There is a clear change in pattern as the sky brightens, and overcomes the noise floor. The mean and $\sigma$ values are given in the title of each flat. The image has been flipped and rotated compared to the images in Figure \ref{ALL_flats_science_moved_reg} to match the telescope correction. This is the reason for the blue pixels at each corner of the image. The PA vectors are also plotted. The $\sigma$ value refers to the random and systematic error over the FOV of interest in Figure \ref{ALL_flats_science_moved_reg} for the 3 $\times$ 3 pixel area indicated by the rectangle on the final image.}
  \label{DOLP_science_flat_fields_moved_crop}
\end{figure}

The results for the DOLP are found in Figure \ref{DOLP_science_flat_fields_moved_crop}. Some of the patterns visible in the system matrix, in Figure \ref{A_norm_flat_moved_section_4x4_science_crop}, appear in the same location for the gain coefficients on RP2, TP1, and TP2. Overall, a flat region of linear polarisation of this section of sky is measured with a minor gradient from the bottom-left-corner to the top-right-corner. The largest value for $\sigma$ per frame is about 1\% for frames with higher SNR.

\begin{figure}[h]
  \centering
  \includegraphics[scale=0.58]{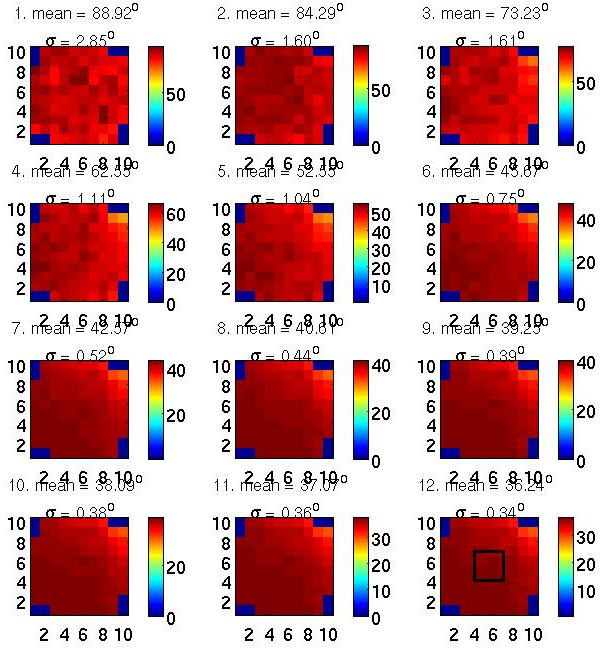}
  \caption{The polarisation angle as a function of time, and field, are plotted. These images are produced using a registration method by science data points. The PA is measured as the Sun rises; the first was summed and averaged for 48 seconds, images 2 - 11 are averaged over 200 frames (3.3 minutes), and the final image, the brightest sky image, is 134 seconds of data. There is a clear change in pattern as the sky brightens, and overcomes the noise floor.  The mean and $\sigma$ values are given in the title of each flat. The image has been flipped and rotated compared to the images in Figure \ref{ALL_flats_science_moved_reg} to match the telescope correction. This is the reason for the blue pixels at each corner of the image. The $\sigma$ value refers to the random and systematic error over the FOV of interest in Figure \ref{ALL_flats_science_moved_reg} for the 3 $\times$ 3 pixel area indicated by the rectangle on the final image.}
  \label{PA_science_flat_fields_moved_crop}
\end{figure}

The GASP system has been compared to work carried out by \cite{Smith1999} where, if GASP measures a pixel error of $\sim$ 0.3 pixels (propagation of the RMS errors in all 3 channels for 4 $\times$ 4 binning), this could result in $\pm$ 6\% error in the DOLP and DOCP. An error for the PA has not been stated by the authors. There is variation in the pixel-to-pixel values measured by the GASP system that are comparable to this result in the extended FOV, outside of that in Figure \ref{ALL_flats_science_moved_reg}.

\begin{figure}[h]
  \centering
  \includegraphics[scale=0.58]{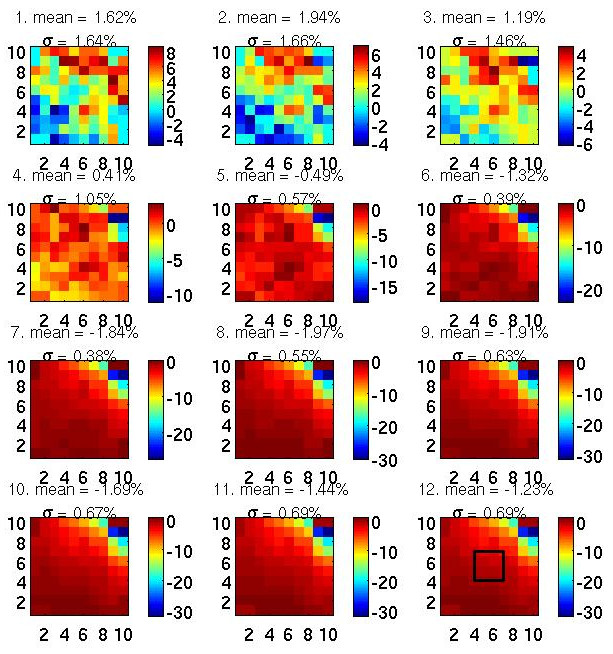}
  \caption{The degree of circular polarisation as a function of time, and field, are plotted. These images are produced using a registration method by science data points. The DOCP is measured as the Sun rises; the first was summed and averaged for 48 seconds, images 2 - 11 are averaged over 200 frames (3.3 minutes), and the final image, the brightest sky image, is 134 seconds of data. There is a clear change in pattern as the sky brightens, and overcomes the noise floor. The mean and $\sigma$ values are given in the title of each flat. The image has been flipped and rotated compared to the images in Figure \ref{ALL_flats_science_moved_reg} to match the telescope correction. The $\sigma$ value refers to the random and systematic error over the FOV of interest in Figure \ref{ALL_flats_science_moved_reg} for the 3 $\times$ 3 pixel area indicated by the rectangle on the final image.}
  \label{DOCP_science_flat_fields_moved_crop}
\end{figure}

\clearpage

A similar trend in the pattern for the polarisation angle (PA) is found in Figure \ref{PA_science_flat_fields_moved_crop}. The mean and $\sigma$ values are taken from the 3 $\times$ 3 pixel area highlighted in the last frame in Figure \ref{DOLP_science_flat_fields_moved_crop}. A low error value has been found in a reasonably flat section of the field, comparable to the same area in the polarimetric measurements for the DOLP and DOCP. The same areas of the FOV as those for DOLP show a gradual gradient in the pattern, particularly from the bottom left to top right corner. This area is also present in the registered TP2 images and appears in the calibration matrix. It is noted that the polarimetric results have been flipped and rotated to account for the telescope orientation on the sky. This has not been performed for the calibration measurements. The variation in pattern is a close match to the image of the TP1 gain coefficient in Figure \ref{A_norm_flat_moved_section_4x4_science_crop}. A comparison is also made between the gain coefficient for the polarisation angle and RP2. This makes sense as the gain coefficient for RP2 and TP1 are used to calculate the values for Stokes Q and U respectively when calculating the PA.
\par
The first 6 frames show a combination of random and systematic error in the results for the DOLP. This has not changed when measuring the PA and the conclusions are the same in terms of the calibrated gain of the system. Table \ref{tab:pol_info_PA_expected} shows that over a 35 minute period the expected change in angle of the Sun with respect to the Earth is $\sim$ 6.3$^{\circ}$.
\par
The pattern in the images for the DOCP in Figure \ref{DOCP_science_flat_fields_moved_crop} also show systematic variation. The highlighted area in the final frame measures the DOCP at 0.7 $\pm$ 0.3\% and appears relatively flat, which indicates that GASP is measuring a low level of circular polarisation, almost zero, and a reasonable $\sigma$ variation. This area is used as outside the total blue section there is an indication that the field is less flat (where the colour map changes from blue to red, left-to-right). There is a large increase in DOCP and the visible shape in the bottom right corner is altered compared to the left-hand-side. This corner is showing quite a large variation in optical distortion not corrected by geometrical registration. It is also noted that this pattern exactly matches that of the TP2 gain coefficient in Figure \ref{A_norm_flat_moved_section_4x4_science_crop}, which measures the Stokes V parameter. This pattern is also found in the raw image and flat-field residuals for TP2 images.

\begin{savenotes}
\begin{table}[h]
\begin{center}
\caption{Mean polarimetric results for Zenith flat-field data using a pixel-by-pixel analysis for the 3 $\times$ 3 pixel area in the final image of each polarimetric measurement. The $\Delta$PA values are also given.}
\label{tab:pol_info_pixel}
\resizebox{1.0\columnwidth}{!}{%
\begin{tabular}{l l l l l l l l l l l l l}
\hline\noalign{\smallskip}
\textbf{Flat-field} & \textbf{1} & \textbf{2} & \textbf{3} & \textbf{4} & \textbf{5} & \textbf{6} & \textbf{7} & \textbf{8} & \textbf{9} & \textbf{10} & \textbf{11} & \textbf{12} \\ 
\noalign{\smallskip}\hline\noalign{\smallskip}
Time (mins) & 0.47 & 2.45 & 5.75 & 9.05 & 12.35 & 15.65 & 18.95 & 22.25 & 25.55 & 28.85 & 32.15 & 35.55 \\ 
\noalign{\smallskip}\hline\noalign{\smallskip}
DOLP (\%) & 35.63 & 35.17 & 31.71 & 35.67 & 44.36 & 57.40 & 69.63 & 79.23 & 84.10 & 86.01 & 87.93 & 90.02 \\ 
\noalign{\smallskip}\hline\noalign{\smallskip}
DOCP (\%) & 1.62 & 1.94 & 1.19 & 0.41 & -0.49 & -1.32 & -1.84 & -1.97 & -1.91 & -1.69 & -1.44 & -1.23 \\ 
\noalign{\smallskip}\hline\noalign{\smallskip}
PA ($^{\circ}$) & 88.92 & 84.29 & 73.23 & 62.55 & 52.55 & 45.67 & 42.57 & 40.61 & 39.25 & 38.09 & 37.07 & 36.24 \\ 
\noalign{\smallskip}\hline\noalign{\smallskip}
$\Delta$PA ($^{\circ}$) & \centering{-} & 4.63 & 11.06 & 10.68 & 10.00 & 6.88 & 3.10 & 1.96 & 1.36 & 1.16 & 1.02 & 0.83 \\ 
\hline\noalign{\smallskip}
\end{tabular}}
\end{center}
\end{table}
\end{savenotes}

The results in Table \ref{tab:pol_info_pixel} are measurements of the mean in the highlighted 3 $\times$ 3 pixel area in the final image of each polarimetric measurement. A variation in the polarisation angle of $\sim$ 52$^{\circ}$ over the course of the observation. Realistically, this value should be measured for when there are less influences from random variations as a result of detector noise. These initial measurements of DOLP are mostly at the noise level, which will cause the polarisation to be overestimated. There appears to be a significant increase in the light level at frames 5 - 6, which gives a change in the PA of $\sim$ 7$^{\circ}$. The lowest frame-to-frame variation in PA observed is 0.83 $\pm$ 0.10$^{\circ}$\footnote{Errors quoted are based on the Poissonian statistics for the 3 $\times$ 3 pixel area}. The $\Delta$PA value is decreasing over time, approaching 0.83$^{\circ}$, compared with an expected variation of 0.71$^{\circ}$. It is possible that if more data was recorded the $\Delta$PA would terminate at this value. Evidence for a dependence on a high level of counts per pixel is clear, however, more data is required over a longer observing period.
\par
The value for the DOLP increases in value between frames 4 - 9, where there is a significant change in light level. However, the $\Delta$ value, frame-to-frame, for the DOLP reaches a terminal value between frames 9 and 10. From frame 9 onwards, the $\Delta$DOLP is 2\%, though with limited data it is difficult to comment on this trend. The change in DOLP over time, and as a function of PA shows good agreement with the Rayleigh Sky Model, however, the experimental measurements indicate that something in addition to Rayleigh scatter is detected by GASP. The measured DOCP value changes by 1 - 2\% around the expected value of 0\%, eventually converging at 0\%. This change is slow, by 0.2\% in the final frames reaching a value of \textminus 1.23 $\pm$ 0.3\%. Deviations from the Rayleigh Sky model have already been discussed, and given the nature of these observations, it is unlikely that GASP has measured pure Rayleigh scatter given the poor observing conditions which include poor seeing, cloud cover and unknown scatter from other local sources or contaminants. However, in all cases each result indicates that GASP is measuring a field dependent, changing polarimetric signal.

\section{Conclusion}

The use of the ECM for an APD detector configuration indicate that GASP works well as a division of amplitude polarimeter. This was concluded as a comparable system matrix to the theoretical PSA was obtained and tested on linearly polarised light. The ECM measured absolute values for the DOLP, DOCP, and PA, with variation in these values as low as 0.2\% and 0.1$^{\circ}$ (in the absence of systematic error), respectively. This experiment also showed a systematic error in the Stokes parameters, which indicates that GASP was measuring a possible mechanical movement, or change in temperature, in the system.
\par
Calibrating GASP using EMCCDs was a more difficult process, given the optical system that was needed to adapt the instrument for observing on the 200 inch at Palomar. The main difficultly was due to image registration when post-processing the data, however, a simple laboratory experiment can verify that these errors are low ($<$ \%1), but optical distortion (after image registration) remains at the edge of the field. These verification results show that it was possible to calibrate GASP, with error; the calibration was further tested on a set of data recorded while the telescope was pointed at the Zenith. It was found that there was little variability in the spatial polarimetric signal across the FOV and that, in spite of registration errors, it was possible to reduce a value for the polarisation angle of the Sun based on its geometric position relative to the Earth. These results are comparable to those predicted by the Rayleigh Sky Model, however, further observations, for extended periods, and higher SNR, are required to investigate/verify the results found from this experiment.

\bibliographystyle{apalike}      
\bibliography{longtitles,gk_bibliography}

\end{document}